\documentclass[5p,times,fleqn]{elsarticle}
\usepackage{ecrc}
\makeatletter
\let\origpprintTitle\ps@pprintTitle
\def\ps@pprintTitle{%
  \origpprintTitle
  \def\@oddfoot{\parbox{\columnwidth}{\footnotesize\itshape
    2213-8463\/ \copyright\ 2026 The Authors. Published by Elsevier Ltd.
    This is an open access article under the CC BY-NC-ND license
    (\url{http://creativecommons.org/licenses/by-nc-nd/4.0/})\\
    Peer-review under responsibility of the scientific committee of the NAMRI/SME.}}%
  \let\@evenfoot\@oddfoot}
\makeatother
\usepackage{amsmath}
\usepackage{float}

\volume{00}

\firstpage{1}

\journalname{Manufacturing Letters}

\runauth{S. Li et al.}

\jid{MFGLET}

\usepackage{amssymb}
\usepackage{ecrc}
\usepackage{amsmath}
\usepackage{mathtools}
\usepackage{booktabs}
\usepackage{algorithm}
\usepackage{algorithmic}

\AtBeginDocument{
  \setlength{\abovedisplayskip}{5pt}
  \setlength{\belowdisplayskip}{5pt}
  \setlength{\abovedisplayshortskip}{1pt}
  \setlength{\belowdisplayshortskip}{1pt}
}

\usepackage{siunitx}

\usepackage[figuresright]{rotating}
\usepackage{geometry}
\usepackage[bookmarks=false]{hyperref}
    \hypersetup{colorlinks,
      linkcolor=blue,
      citecolor=blue,
      urlcolor=blue}

\usepackage{titlesec}
\titlespacing*{\section}{0pt}{8pt plus 2pt minus 2pt}{4pt plus 2pt minus 2pt}
\titlespacing*{\subsection}{0pt}{6pt plus 2pt minus 2pt}{3pt plus 2pt minus 2pt}
\titlespacing*{\subsubsection}{0pt}{4pt plus 2pt minus 2pt}{2pt plus 2pt minus 2pt}

\begin{document}

\begin{frontmatter}



\dochead{54th SME North American Manufacturing Research Conference (NAMRC 54, 2026)}%

\title{\textbf{Curriculum-Based Soft Actor-Critic for Multi-Section R2R Tension Control}}


\author[a]{Shihao Li} 
\author[a]{Jiachen Li}
\author[a]{Christopher Martin}
\author[a]{Zijun Chen}
\author[a]{Dongmei Chen}
\author[a]{Wei Li\corref{cor1}}
\ead{weiwli@austin.utexas.edu}

\address[a]{Walker Department of Mechanical Engineering, University of Texas at Austin, Austin, TX 78712, USA}
\cortext[cor1]{Corresponding author. Tel.: +1-512-471-7174}

\begin{abstract}
Precise tension control in roll-to-roll (R2R) manufacturing systems remains challenging under varying operating conditions and process uncertainties. While they perform well for systems with accurate models and small disturbances, model-based controllers face limitations when operating conditions deviate from design points or when process uncertainties are large. This paper presents a curriculum-based Soft Actor-Critic (SAC) approach as an alternative to R2R control under uncertainty. The method learns control policies through progressive exposure to various operating conditions without requiring system-specific parameter tuning. We train SAC with a three-phase curriculum with gradual expansion of reference trajectory diversity, from a narrow ranges (\SI{27}{\newton}--\SI{33}{\newton}) to full operational envelope (\SI{20}{\newton}--\SI{40}{\newton}), enabling the policy to generalize across the complete operational range. Evaluation on a three-section R2R system shows that the learned policy maintains tight tension tracking under nominal conditions and handles large disturbances (\SI{20}{\newton}$\rightarrow$\SI{40}{\newton} step changes) well with a single control policy. The proposed approach offers an alternative to traditional model-based methods for applications where operating conditions vary and process uncertainties are significant, typical for R2R manufacturing of flexible devices.
\end{abstract}


\end{frontmatter}

% \linenumbers


\begin{nomenclature}

\begin{deflist}[$\omega_i$]
\defitem{\textbf{Nomenclature}}\defterm{}
\defitem{}\defterm{}
\defitem{$E$}\defterm{Modulus of web (N/m$^2$)}
\defitem{$A$}\defterm{Cross-sectional area of web (m$^2$)}
\defitem{$R_i$}\defterm{Radius of roller $i$ (m)}
\defitem{$J_i$}\defterm{Inertia of roller $i$ (kg-m$^2$)}
\defitem{$f_i$}\defterm{Friction coefficient motor $i$ (N-m-s-rad$^{-1}$)}
\defitem{$L_i$}\defterm{Length of web section $i$ (m)}
\defitem{$b_i$}\defterm{Disturbance coefficient motor $i$ (s-m$^{-1}$-kg$^{-1}$)}
\defitem{$T_i$}\defterm{Tension in web section $i$ (N)}
\defitem{$v_i$}\defterm{Linear velocity over roller $i$ (m-s$^{-1}$)}
\defitem{$v_0$}\defterm{Unwinding velocity (m-s$^{-1}$)}
\defitem{$\omega_i$}\defterm{Rotational velocity of roller $i$ (rad-s$^{-1}$)}
\defitem{$u_i$}\defterm{Motor torque applied to motor $i$ (N-m)}
\end{deflist}
\end{nomenclature}

%\enlargethispage{-7mm}

\section{Introduction}
\label{sec:introduction}

Roll-to-roll (R2R) manufacturing is essential for high-throughput production of flexible electronics, photovoltaics, batteries, and advanced materials \cite{martin2025review,krebs2009roll,abbel2018roll}. These systems transport continuous webs through multiple processing stages at speeds ranging from meters to hundreds of meters per minute, demanding precise tension and velocity control to prevent wrinkles, telescoping, and registration errors \cite{ma2025factors,chen2022nonlinear}. Industrial R2R lines face significant operational diversity, such as varying substrate elastic properties, startup and shutdown transients, tension adjustments for different coating thicknesses, and processing temperature variations. Existing R2R process control relies primarily on model-based optimal control methods including Linear Quadratic Regulators (LQR), H-infinity robust control, and Model Predictive Control (MPC) \cite{martin2025review,chen2023control,zhao2023realtime}. Recent advances include derivative-free trajectory bundle methods that achieve constraint-guaranteed MPC without gradient computation \cite{li2025adaptive} and LLM-assisted frameworks that automate controller commissioning and adaptation \cite{li2025llm}. While these approaches advance the state of the art, model-based methods still face challenges when operating conditions deviate significantly from design points and process uncertainties are large. For example, classical adaptive control and gain-scheduling approaches have been explored for web handling in time-varying R2R systems \cite{raul2015design}; however, these approaches still require fairly accurate analytical models and manual tuning for each operating regime.

Recent work in machine learning and AI has demonstrated the potential of reinforcement learning (RL) for manufacturing process control. RL has been applied to chemical 
process optimization \cite{nian2020review}, CNC machining parameter tuning, and robotic assembly tasks, with deep RL methods showing particular promise for complex, high-dimensional control problems \cite{panzer2022deep}. Despite growing interest in RL for manufacturing process control, current approaches typically train on fixed task distributions without progressive difficulty scheduling, limiting their ability to generalize across wide operational envelopes. 

In this research, we propose a Soft Actor-Critic (SAC)  with Curriculum Learning approach to R2R process control under significant uncertain operating conditions.  SAC \cite{haarnoja2018softa,haarnoja2018softb,zeng2024multi,eysenbach2021maximum}learns control policies from interaction data without assuming specific system dynamics.  It also employs deep neural networks to represent nonlinear policies and enable coordinated control strategies capable of handling strong inter-span coupling, where tension changes in one section propagate to adjacent spans in R2R systems \cite{chen2023control}. Our method trains a single policy to handle both nominal tracking and disturbance rejection without requiring scenario-specific tuning. To handle diverse operating conditions, we employ a curriculum learning strategy \cite{bengio2009curriculum, narvekar2020curriculum} to expose the policy to increasing variation during training, so that it learns to maintain nominal tracking performance while developing responses to large disturbances. This progressive exposure enables more stable learning than naive full-range training and can produce policies that generalize across both nominal conditions and disturbance scenarios---a critical capability for industrial R2R systems requiring operational flexibility. We demonstrate the proposed methodology by comparing its performance against that of traditional model-based control approaches for an R2R process with significant process uncertainties and varying operating conditions. 

\section{Methodology}
\label{sec:background}

\subsection{Roll-to-Roll System Dynamics}
\label{subsec:r2r_dynamics}

\begin{figure}[htbp]
    \centering
    \includegraphics[width=0.4\textwidth]{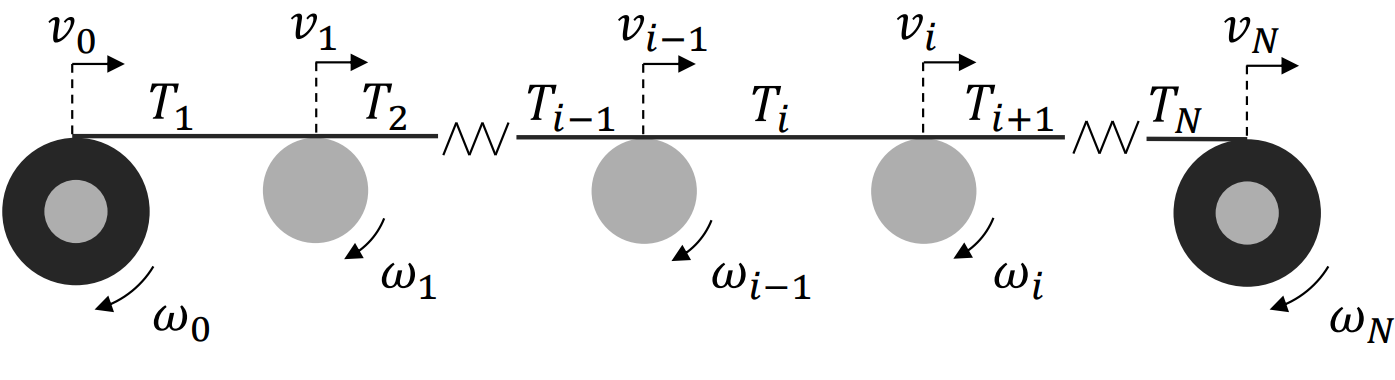}
    \caption{Schematic of a simplified R2R line}
    \label{fig:r2r_schematic}
\end{figure}

A typical R2R manufacturing line consists of $N$ active rollers separated by substrate spans. The system state at time $t$ is characterized by web tensions $\mathbf{T}(t) = [T_1, T_2, \ldots, T_N]^T$ and velocities $\mathbf{v}(t) = [v_1, v_2, \ldots, v_N]^T$. Control inputs are motor torques $\mathbf{u}(t) = [u_1, u_2, \ldots, u_M]^T$ applied to each roller. Figure~\ref{fig:r2r_schematic} illustrates a multi-section R2R system, which consists of an unwinder (roller 0), $N$ processing rollers with rotational 
velocities $\omega_i$, and a rewinder (roller N). Web tensions $T_i$ develop 
in elastic spans (indicated by spring symbols) between adjacent rollers, 
with linear velocities $v_i$ maintained at each roller.

The continuous-time dynamics for a processing section $i$ are given by \cite{chen2023control}:

\begin{equation}
\dot{T}_i = \frac{EA}{L_i}(v_i - v_{i-1}) + \frac{1}{L_i}(v_{i-1}T_{i-1} - v_i T_i)
\label{eq:tension_dynamics}
\end{equation}

\begin{equation}
\dot{v}_i = \frac{R^2}{J}(T_{i+1} - T_i) - \frac{f_b}{J}v_i + \frac{R}{J}u_i
\label{eq:velocity_dynamics}
\end{equation}

where $EA$ is web stiffness, $L_i$ is the length of span $i$, $R$ is roller radius, $J$ is roller inertia, and $f_b$ is friction coefficient. These equations exhibit strong coupling between adjacent sections through the tension terms $T_{i-1}$ and $T_{i+1}$, making the system a complex multi-input multi-output (MIMO) control problem. In this work, we assume equal web span lengths ($L_i = L$ for all $i$) for simplicity, though the methodology extends to non-uniform configurations.

\subsection{Soft Actor-Critic}
\label{subsec:sac}
We formulate R2R tension control as a Markov Decision Process (MDP) defined by the tuple $(\mathcal{S}, \mathcal{A}, \mathcal{P}, \mathcal{R}, \gamma)$, where $\mathcal{S}$ and $\mathcal{A}$ are continuous state and action spaces, $\mathcal{P}(s'|s,a)$ is the transition probability, $\mathcal{R}(s,a)$ is a reward function, and $\gamma \in [0,1)$ is a discount factor. The specific MDP instantiation for our problem is detailed in Section~\ref{subsec:mdp_formulation}.

We solve this MDP using Soft Actor-Critic (SAC)~\cite{haarnoja2018softa,haarnoja2018softb}, an off-policy deep reinforcement learning algorithm. SAC is model-free: it learns an optimal policy from sampled transitions $(s_t, a_t, r_t, s_{t+1})$ obtained through environment interaction, without requiring explicit knowledge of the transition dynamics $\mathcal{P}$. This is advantageous for R2R systems where accurate analytical models may be difficult to obtain due to material property variations.

SAC's distinguishing feature is entropy-regularized optimization. Rather than maximizing expected cumulative reward alone, SAC maximizes a combined objective that encourages stochastic, exploratory policies:

\begin{equation}
J(\pi) = \mathbb{E}_{\tau \sim \pi} \left[ \sum_{t=0}^{\infty} \gamma^t \left( \mathcal{R}(s_t, a_t) + \alpha \mathcal{H}(\pi(\cdot|s_t)) \right) \right]
\label{eq:sac_objective}
\end{equation}

where $\mathcal{H}(\pi(\cdot|s_t)) = \mathbb{E}_{a \sim \pi(\cdot|s_t)}[-\log \pi(a|s_t)]$ is the policy entropy at state $s_t$ and $\alpha > 0$ is an automatically tuned temperature parameter controlling the exploration--exploitation trade-off. This entropy regularization improves robustness to disturbances and model uncertainties~\cite{eysenbach2021maximum}---a valuable property for R2R systems with material property variations and unmeasured disturbances. SAC maintains a stochastic actor network $\pi_\phi(a|s)$ and two critic networks $Q_{\theta_1}(s,a)$, $Q_{\theta_2}(s,a)$ to mitigate value overestimation; architectural details are provided in Section~\ref{subsec:network_architecture}..

\subsection{Formulation of Markov Decision Process}
\label{subsec:mdp_formulation}

We formulate the R2R tension control problem as a single-agent continuous control Markov Decision Process (MDP) where one centralized policy controls all roller torques. This centralized architecture handles inter-span coupling by learning coordinated control strategies across all sections.

The transition dynamics $\mathcal{P}(s'|s,a)$ of this MDP are implicitly 
defined by the R2R system equations (Eqs.~\ref{eq:tension_dynamics}--\ref{eq:velocity_dynamics}), 
discretized via Euler integration with timestep $dt = 0.01$\,s, plus 
additive Gaussian process noise ($\sigma = 0.05$) on control inputs. 
As a model-free algorithm, SAC does not require an explicit representation 
of $\mathcal{P}$; instead, it learns directly from sampled transitions 
$(s_t, a_t, r_t, s_{t+1})$ collected through environment interaction. 
This bypasses the need for system identification or analytical transition 
models---a practical advantage for R2R systems where accurate dynamic 
models may be difficult to obtain due to material property variations 
and complex web--roller interactions.

\subsubsection{State Space}

Unlike traditional control approaches that use only current state deviations, we design a rich observation space to provide the policy with comprehensive system information. For an N-section R2R system, the observation vector at time $t$ consists of:

\begin{equation}
\begin{split}
s_t = [&\mathbf{T}_{norm}, \mathbf{v}_{norm}, \mathbf{T}^{ref}_{norm}, \mathbf{v}^{ref}_{norm}, \\
&\mathbf{e}_T, \mathbf{e}_v, \mathbf{u}_{t-1,norm}, \phi_t]^T \in \mathbb{R}^{7N+1}
\end{split}
\label{eq:state_space}
\end{equation}

where each vector component dimension corresponds to the number of sections N. The components are: $\mathbf{T}_{norm} \in \mathbb{R}^N$ (normalized current tensions for each section), $\mathbf{v}_{norm} \in \mathbb{R}^N$ (normalized velocities), $\mathbf{T}^{ref}_{norm}, \mathbf{v}^{ref}_{norm} \in \mathbb{R}^N$ (normalized references), $\mathbf{e}_T, \mathbf{e}_v \in \mathbb{R}^N$ (tracking errors), $\mathbf{u}_{t-1,norm} \in \mathbb{R}^N$ (normalized previous actions), and $\phi_t \in [0,1]$ (normalized episode progress, scalar). This yields $7N + 1$ total observation dimensions. All state variables are normalized to facilitate neural network training. Including tracking errors helps the policy focus on the control objective. The previous action $\mathbf{u}_{t-1}$ provides temporal context, enabling the policy to learn smooth control strategies and avoid abrupt changes. The episode progress $\phi_t$ allows the policy to distinguish early-episode exploration from late-episode fine-tuning behavior, which can be valuable for developing stable startup and steady-state control strategies. This rich observation space allows SAC to learn more nuanced control policies compared to traditional state-feedback controllers that use only current state deviations.

\subsubsection{Action Space}

The action at time $t$ consists of motor torques for all N rollers:

\begin{equation}
a_t = [u_1, u_2, \ldots, u_N]^T \in [-1, 1]^N
\label{eq:action_space}
\end{equation}

Actions are continuous and normalized to the range $[-1, 1]$. SAC's policy first samples unbounded actions from a Gaussian distribution, then applies a $\tanh$ transformation to map them into the bounded interval $[-1, 1]$. This "squashing" transformation ensures actions respect physical actuator limits while maintaining the benefits of Gaussian exploration during training. These normalized actions are scaled to physical torques as $u_{physical,i} = a_i \cdot u_{scale}$ where $u_{scale}$ represents the maximum allowable motor torque.

To encourage smooth control and mitigate high-frequency oscillations common in learned controllers, we apply an exponential moving average (EMA) filter to the actions:

\begin{equation}
\tilde{u}_t = \beta u_t + (1-\beta) \tilde{u}_{t-1}
\label{eq:action_smoothing}
\end{equation}

where $\beta$ is the smoothing coefficient and $\tilde{u}_t$ is the smoothed action applied to the system. This filtering is integrated into the environment dynamics rather than the policy network, ensuring smooth control even during exploration.

\subsubsection{Reward Function}

The reward function plays a critical role in shaping the learned behavior. We design a multi-component reward that balances multiple objectives:

\begin{equation}
\begin{split}
r_t = \lambda \Big[ &-w_T \cdot \text{MSE}(\mathbf{T}, \mathbf{T}^{ref}) - w_v \cdot \text{MSE}(\mathbf{v}, \mathbf{v}^{ref}) \\
&- w_c \cdot ||\mathbf{u}||^2 - w_s \cdot ||\mathbf{u}_t - \mathbf{u}_{t-1}||^2 \\
&- w_{viol} \cdot V(\mathbf{T}) + w_{succ} \cdot I_{success} \Big]
\end{split}
\label{eq:reward_function}
\end{equation}

where the components are:

\begin{itemize}
    \item \textbf{Tension tracking:} $-w_T \cdot \text{MSE}(\mathbf{T}, \mathbf{T}^{ref})$ penalizes mean squared error in tension tracking

    \item \textbf{Velocity tracking:} $-w_v \cdot \text{MSE}(\mathbf{v}, \mathbf{v}^{ref})$ penalizes velocity deviations

    \item \textbf{Control magnitude:} $-w_c \cdot ||\mathbf{u}||^2$ discourages excessive control effort

    \item \textbf{Control smoothness:} $-w_s \cdot ||\mathbf{u}_t - \mathbf{u}_{t-1}||^2$ penalizes rapid control changes

    \item \textbf{Constraint violations:} $-w_{viol} \cdot V(\mathbf{T})$ heavily penalizes tension violations outside safe operational bounds

    \item \textbf{Success bonus:} $w_{succ} \cdot I_{success}$ provides sparse reward when tracking performance meets specified thresholds
\end{itemize}

All components are scaled by $\lambda = 0.01$ to keep rewards in an 
approximate $[-10, +1]$ range, which we empirically found to improve 
training stability.

\subsection{Neural Network Architecture}
\label{subsec:network_architecture}

Our SAC implementation uses three neural networks: one actor (policy) network and two critic (Q-value) networks. All networks are multilayer perceptrons (MLPs) with shared architectural design principles. During operation, the actor network takes the current observation $s_t$ as input and outputs an action $a_t$ (motor torques). This action is executed on the R2R system, producing the next state $s_{t+1}$ and reward $r_t$ based on Equation~\ref{eq:reward_function}. The experience tuple $(s_t, a_t, r_t, s_{t+1})$ is stored in a replay buffer. During training, the algorithm randomly samples batches of these experience tuples from the buffer to update all three networks: the critics learn to predict cumulative future rewards $Q(s,a)$ by minimizing temporal difference errors, while the actor learns to select actions that maximize the critics' Q-value predictions plus entropy. This off-policy learning approach allows SAC to reuse past experiences, learning from hundreds of thousands of interactions with the simulated R2R system.

\textbf{Actor Network:} The policy network $\pi_\phi(a|s)$ takes the $(7N+1)$-dimensional observation vector as input and outputs a stochastic policy over the N-dimensional action space. The architecture consists of three fully-connected hidden layers, each with 256 units, followed by two output heads that predict the mean $\mu(s)$ and log-standard deviation $\log \sigma(s)$ of a diagonal Gaussian distribution over actions. We use the Exponential Linear Unit (ELU) activation function \cite{clevert2015fast} for all hidden layers, which we found to provide better gradient flow and training stability compared to Rectified Linear Unit (ReLU) for continuous control tasks. The final action is sampled from the Gaussian and passed through a $\tanh$ squashing function to ensure bounded actions in $[-1, 1]^N$.

\textbf{Critic Networks:} To mitigate overestimation bias, SAC employs two Q-networks $Q_{\theta_1}(s,a)$ and $Q_{\theta_2}(s,a)$ that estimate state-action values \cite{haarnoja2018softb}. Both critics share the same architecture: three hidden layers of 256 units each with ELU activations, taking the concatenated observation-action pair $[s, a]$ as input and outputting a scalar Q-value representing expected cumulative reward. During training, the minimum of the two Q-estimates is used for policy updates, reducing positive bias in value estimation.

\textbf{Target Networks:} Soft target networks $Q_{\bar{\theta}_1}$ and $Q_{\bar{\theta}_2}$ are maintained as slowly-updating copies of the critic networks to stabilize training, updated via Polyak averaging with coefficient $\tau = 0.005$ \cite{lillicrap2015continuous}.

\textbf{Network Initialization:} All network weights are initialized using orthogonal initialization \cite{hu2020provable}, which has been shown to improve learning speed for reinforcement learning tasks.

\subsection{Training Procedure}
\label{subsec:training_procedure}

\textbf{Training Data Source:} Unlike supervised learning methods that require pre-collected datasets, reinforcement learning generates its own training data through direct interaction with the environment. The SAC agent learns by controlling a high-fidelity simulation of the R2R system. The simulation implements the nonlinear dynamics (Equations~\ref{eq:tension_dynamics}--\ref{eq:velocity_dynamics}) and provides the agent with observations, rewards, and state transitions at each timestep. During training, the agent accumulates experience tuples $(s_t, a_t, r_t, s_{t+1})$ by trying different control actions and observing their effects. This simulation-based approach eliminates the need for expensive and time-consuming physical experiments during training, while still producing controllers that can be transferred to real hardware after domain adaptation.

We train the SAC agent using the entropy-regularized RL objective with automatic temperature tuning \cite{haarnoja2018softa}. Unlike standard RL training that exposes the agent to the full task complexity from the start, we employ a three-phase curriculum learning strategy that increases task difficulty. This curriculum enables more stable learning and better generalization than naive domain randomization \cite{bengio2009curriculum}.

\subsubsection{Curriculum Learning Strategy}

The key insight behind curriculum learning is that complex control tasks are easier to learn when the agent is first trained on simplified versions before graduating to the full problem. For R2R tension control, we define task difficulty through the range of reference trajectories the agent must track. Our three-phase curriculum expands this range:

\textbf{Phase 1: Foundation (0--40\% of training):} The agent learns basic control capabilities on narrow operating ranges around the nominal setpoint. This constrained regime allows the policy to develop stable control strategies without being overwhelmed by extreme operating conditions. Reference trajectories are sampled from narrow ranges with low probability of step changes to maintain task simplicity.

\textbf{Phase 2: Expansion (40--80\% of training):} As the policy demonstrates competence on easy scenarios (indicated by stabilizing evaluation rewards), the curriculum transitions to moderate difficulty. Operating ranges expand to intermediate levels, and step change probability increases. This gradual expansion prevents catastrophic forgetting of the foundational skills while pushing the policy to generalize to broader operating conditions.

\textbf{Phase 3: Mastery (80--100\% of training):} The final phase exposes the agent to the full operational envelope. Reference trajectories span the complete range of operating conditions the system is expected to handle, and step changes occur with higher probability. Training on this challenging distribution ensures the policy can handle worst-case scenarios while maintaining the precision learned in earlier phases.

The curriculum phase is determined based on training progress: if $p = \frac{\text{current timestep}}{\text{total timesteps}}$ represents fractional training progress, then Phase 1 applies for $p < 0.4$, Phase 2 for $0.4 \leq p < 0.8$, and Phase 3 for $p \geq 0.8$. This fixed schedule ensures reproducible training progression across different random seeds.

The three-phase structure follows 
established curriculum learning practice, where tasks are decomposed 
into 2--4 difficulty levels to balance learning stability against 
schedule complexity~\cite{bengio2009curriculum,narvekar2020curriculum}. 
Empirically, we found that a 
40/40/20 split produced more stable convergence than equal 33/33/33 
splits, where premature Phase~2 transitions occurred before Phase~1 
skills had stabilized. The reference ranges were chosen to create 
meaningful difficulty increments: Phase~1 covers $\pm$10\% around the 
nominal 30\,N setpoint (27\,N--33\,N), Phase~2 extends to $\pm$17\% 
(25\,N--35\,N), and Phase~3 spans the full operational envelope 
(20\,N--40\,N). This geometric-like expansion ensures each phase 
introduces a substantial but manageable increase in task diversity.

\subsubsection{Standard SAC Training Loop}

Within each curriculum phase, we follow the standard SAC training procedure:

\textbf{Exploration Phase:} For the first 10,000 timesteps, the agent executes random actions sampled uniformly from $[-1, 1]^N$ to populate the replay buffer with diverse experiences. This warm-up period helps establish a broad coverage of the state-action space before gradient-based learning begins.

\textbf{Learning Phase:} After the warm-up period, the agent alternates between environment interaction and policy updates. At each timestep, the agent:
\begin{enumerate}
    \item Samples an action from the current policy $a_t \sim \pi_\phi(\cdot|s_t)$
    \item Executes the action and observes the next state, reward, and termination signal
    \item Stores the transition $(s_t, a_t, r_t, s_{t+1}, done_t)$ in the replay buffer (capacity: 1,000,000 transitions)
    \item Samples a mini-batch of 256 transitions uniformly from the replay buffer
    \item Performs one gradient descent step on the critic loss and one step on the actor loss
\end{enumerate}

\textbf{Key Hyperparameters:}
\begin{itemize}
    \item \textbf{Learning rate:} $3 \times 10^{-4}$ for all networks (Adam optimizer \cite{kingma2014adam})
    \item \textbf{Discount factor:} $\gamma = 0.99$ (long-term planning)
    \item \textbf{Soft update coefficient:} $\tau = 0.005$
    \item \textbf{Batch size:} 256 transitions
    \item \textbf{Training frequency:} One gradient step per environment step
    \item \textbf{Entropy temperature:} Automatically tuned to target entropy $-\dim(\mathcal{A}) = -N$
\end{itemize}

\textbf{Episode Structure:} Each training episode consists of 500 timesteps (\SI{5}{\second} of simulation time with $dt = \SI{0.01}{\second}$), representing one complete control trial from start to finish. After each episode, the simulation resets to new initial conditions with randomized reference trajectories sampled from the current curriculum phase. Training across approximately 1,000 episodes exposes the agent to diverse operating conditions (different tension references, velocity setpoints, and disturbance patterns), enabling generalization across the full operational envelope rather than overfitting to a single scenario.

\textbf{Evaluation Protocol:} Every 5,000 training steps, we evaluate the current policy deterministically (using mean actions without sampling) on 5 independent episodes using the current curriculum phase's task distribution. The best model is selected based on mean evaluation reward and saved for final comparison with baselines.

\textbf{Convergence Criteria:} Training proceeds for a total of 500,000 timesteps (approximately 1,000 episodes) for each random seed, which we found sufficient for convergence based on the stabilization of evaluation rewards across all seeds. The best-performing checkpoint based on mean evaluation reward across all three seeds appeared at 440,000 timesteps (during Phase 3, late in training), achieving a mean evaluation reward of 4.87 with narrow confidence intervals. This late-stage peak performance demonstrates that the curriculum enabled the policy to master challenging scenarios while maintaining the foundational skills acquired in earlier phases.

\section{Case Study: Three-Section R2R System}
\label{sec:case_study}

We demonstrate the proposed methodology on a laboratory-scale three-section R2R system ($N=3$). The system consists of an unwinder, three processing rollers with independent motor torque control, and a rewinder (see Figure~\ref{fig:r2r_schematic}). The physical parameters are provided in Table~\ref{tab:system_parameters}.

\begin{table}[h]
\centering
\caption{Physical parameters of the three-section R2R system}
\label{tab:system_parameters}
\begin{tabular}{lc}
\toprule
\textbf{Parameter} & \textbf{Numerical Value} \\
\midrule
Modulus ($E$) & \SI{200}{\mega\pascal} \\
Cross-sectional area ($A$) & $1.2 \times 10^{-5}$ \si{\meter\squared} \\
Roller radius ($R$) & \SI{0.04}{\meter} \\
Roller inertia ($J$) & \SI{0.95}{\kilo\gram\meter\squared} \\
Motor friction ($f_b$) & \SI{10}{\newton\meter\second\per\radian} \\
Web section length ($L$) & \SI{1.0}{\meter} \\
\bottomrule
\end{tabular}
\end{table}

\textbf{Operating envelope:} The nominal operating point is $T_{ref} = \SI{30}{\newton}$ and $v_0 = \SI{0.01}{\meter\per\second}$, representing typical laboratory-scale operation. The system must handle tensions ranging from \SI{20}{\newton} to \SI{40}{\newton} and velocities from \SI{0.005}{\meter\per\second} to \SI{0.015}{\meter\per\second} to accommodate different substrate materials and process requirements.

\textbf{MDP instantiation:} For $N=3$ sections, the state space is $\mathbb{R}^{22}$ ($7 \times 3 + 1 = 22$ dimensions) and the action space is $\mathbb{R}^3$. Normalization constants are: $T_{nominal} = \SI{30}{\newton}$, $T_{range} = \SI{40}{\newton}$ (covering \SI{20}{\newton}--\SI{40}{\newton} with margin), $v_{nominal} = \SI{0.01}{\meter\per\second}$, $v_{range} = \SI{0.02}{\meter\per\second}$. Maximum motor torque is $u_{scale} = \SI{2.0}{\newton\meter}$. Action smoothing coefficient is $\beta = 0.7$.

\textbf{Reward parameters:} $w_T = 100$, $w_v = 1000$, $w_c = 0.1$, $w_s = 0.5$, $w_{viol} = 100$, $w_{succ} = 1.0$, $\lambda = 0.01$. Constraint bounds are [\SI{10}{\newton}, \SI{50}{\newton}]. Success thresholds are $\max_i |T_i - T_i^{ref}| < \SI{0.5}{\newton}$ and $\max_i |v_i - v_i^{ref}| < \SI{0.001}{\meter\per\second}$.

\textbf{Curriculum instantiation:} Phase 1 (0--40\% training): Tensions $\in$ [\SI{27}{\newton}, \SI{33}{\newton}], velocities $\in$ [\SI{0.0095}{\meter\per\second}, \SI{0.0105}{\meter\per\second}], 10\% step change probability. Phase 2 (40--80\%): Tensions $\in$ [\SI{25}{\newton}, \SI{35}{\newton}], velocities $\in$ [\SI{0.009}{\meter\per\second}, \SI{0.011}{\meter\per\second}], 20\% step changes. Phase 3 (80--100\%): Tensions $\in$ [\SI{20}{\newton}, \SI{40}{\newton}], velocities $\in$ [\SI{0.008}{\meter\per\second}, \SI{0.012}{\meter\per\second}], 30\% step changes.

This three-section configuration provides sufficient complexity to demonstrate inter-span coupling effects while remaining computationally tractable for comprehensive multi-seed training.

For testing, we evaluate the trained policy on two specific scenarios not directly encountered during training, enabling assessment of generalization capability:

\textbf{Test Case 1---Nominal Tracking:} Constant tension reference at \SI{30}{\newton} across all sections with steady inlet velocity, representing standard production conditions where traditional controllers excel.

\textbf{Test Case 2---Tension Step Response (Disturbance Rejection):} Section 2 receives a large step change (\SI{20}{\newton} $\rightarrow$ \SI{40}{\newton}) while other spans maintain nominal setpoints, testing inter-span coupling rejection and off-nominal robustness---conditions where traditional controllers struggle.

By evaluating the same learned SAC policy on both scenarios, we assess whether curriculum-based training enables a single policy to handle both nominal tracking and disturbance rejection.

All simulation studies were conducted on the three-section R2R model, whichincorporates two categories of uncertainty: (1)~additive process noise, where Gaussian noise $w_t \sim \mathcal{N}(0, \sigma^2 I)$ with $\sigma = 0.05$ is added to each normalized control input ($\pm$5\% stochastic torque variation), and (2)~operational diversity via the curriculum, which progressively expands reference trajectory ranges across phases as described above. The current study does not randomize physical system parameters (e.g., web stiffness $EA$, roller inertia $J$); extending training via domain randomization over physical parameters is a natural direction for future work.

The control task involves tracking a constant tension reference of \SI{30}{\newton} across all three sections while maintaining steady-state velocity profiles that satisfy continuity constraints. Each evaluation episode runs for 500 timesteps (\SI{5}{\second} of simulated time) with timestep $dt = \SI{0.01}{\second}$.

All three controllers (SAC, MPC, LQR) were evaluated on 10 independent episodes with identical initial conditions and random seeds for fair comparison. For SAC, we used the best checkpoint (285k training steps) selected based on evaluation reward during training. Evaluation metrics include:

\begin{itemize}
    \item \textbf{Tension MAE/RMSE:} Mean absolute error and root mean squared error in tension tracking across all sections
    \item \textbf{Velocity MAE/RMSE:} Mean absolute error and root mean squared error in velocity tracking
    \item \textbf{Episode Reward:} Cumulative reward as defined in Equation \ref{eq:reward_function}
    \item \textbf{Control Smoothness:} Variance of control action changes $\text{Var}(\Delta u)$
\end{itemize}

In this study, MPC and LQR controllers were carefully tuned and operate on the same R2R simulation environment with identical system parameters. The LQR controller uses analytically computed optimal gains based on linearization around the nominal operating point (\SI{30}{\newton} tension, \SI{0.01}{\meter\per\second} velocity), where it achieves optimal performance for that specific condition. The MPC controller solves a finite-horizon quadratic program (QP) at each timestep with a 10-step prediction horizon ($\SI{0.1}{\second}$ lookahead), using carefully selected cost matrices ($Q$, $R$) that balance tracking performance and control effort. Both controllers were iteratively tuned to achieve their best possible performance on the nominal tracking scenario (Test Case 1), ensuring the comparison does not disadvantage model-based methods. This tuning process involved grid search over weighting parameters to minimize tracking error while maintaining control smoothness, representing the performance achievable by experienced control engineers. 

\section{Results and Discussion}
\label{sec:results}

\begin{figure*}[t]
    \centering
    \includegraphics[width=0.95\textwidth]{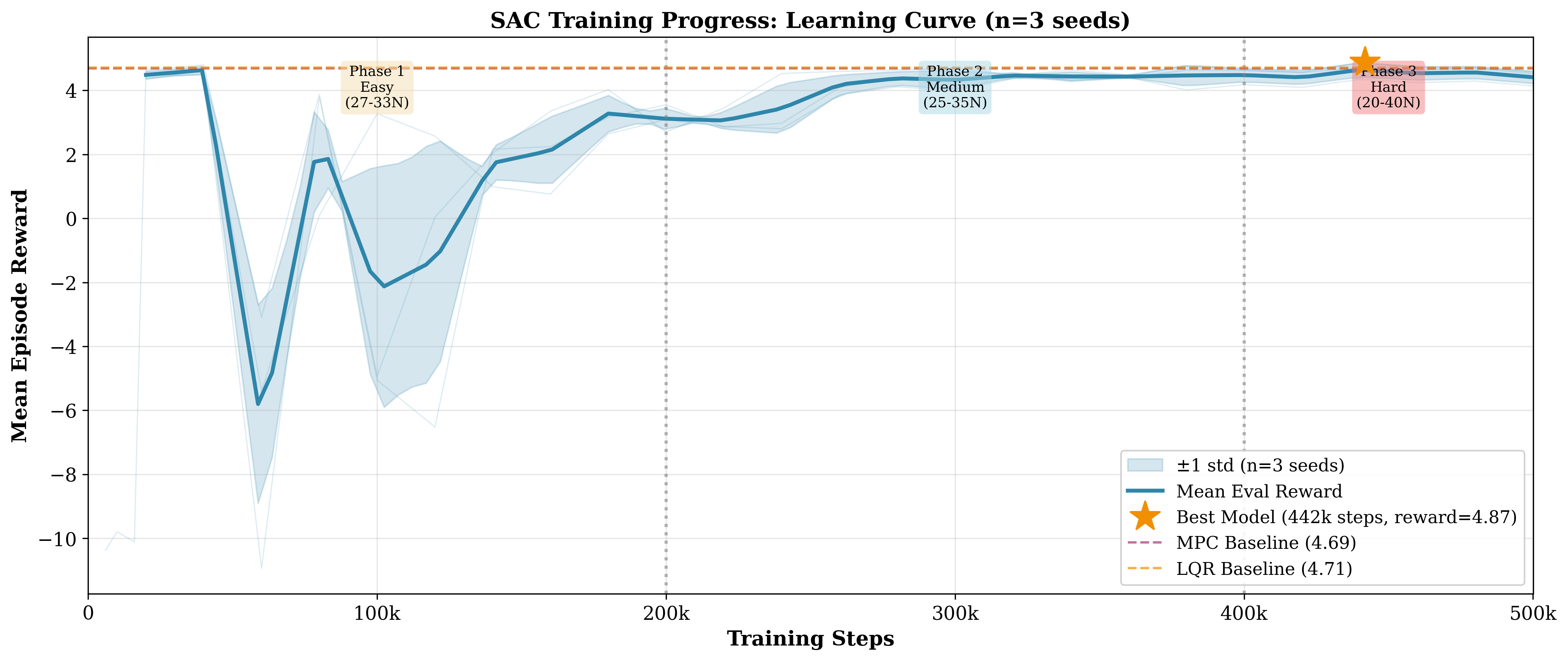}
    \caption{SAC training progress over 500,000 timesteps (approximately 1,000 episodes per seed, where each episode = 500 timesteps of simulated R2R control) with three-phase curriculum learning (n=3 random seeds for statistical rigor, representing 3,000 total episodes). The solid blue line shows mean evaluation reward across all seeds, with shaded region indicating confidence interval (mean $\pm$ 1 std). Faint blue lines show individual seed trajectories. Vertical dotted lines mark curriculum phase transitions at 200k (Phase 1 $\rightarrow$ Phase 2) and 400k steps (Phase 2 $\rightarrow$ Phase 3). Colored boxes indicate tension reference ranges: Phase 1 (\SI{27}{\newton}--\SI{33}{\newton}), Phase 2 (\SI{25}{\newton}--\SI{35}{\newton}), Phase 3 (\SI{20}{\newton}--\SI{40}{\newton}). The orange star marks the best mean performance at 440k steps (reward=4.87). Horizontal dashed lines show MPC (\num{4.69}) and LQR (\num{4.71}) baseline performance for comparison. The narrow confidence intervals demonstrate reproducible learning across random seeds.}
    \label{fig:learning_curve}
\end{figure*}

\subsection{Training Performance}
\label{subsec:training_performance}
Figure~\ref{fig:learning_curve} shows the learning progression of SAC over 500,000 training timesteps with curriculum learning, aggregated across three independent training runs with different random seeds (42, 43, 44) to ensure statistical rigor and reproducibility. The solid blue line represents mean evaluation reward across all seeds, with the shaded region indicating confidence intervals (mean $\pm$ 1 standard deviation). Individual seed trajectories are shown as faint lines. The three vertical dotted lines mark curriculum phase transitions at 200k steps (Phase 1 $\rightarrow$ Phase 2) and 400k steps (Phase 2 $\rightarrow$ Phase 3), with labeled boxes indicating the tension reference ranges for each phase.

The learning curve reveals several important characteristics of curriculum-based training. During Phase 1 (0--200k steps, tension range \SI{27}{\newton}--\SI{33}{\newton}), the agent exhibits high initial volatility (mean rewards dropping as low as $-10$ during early exploration) before converging to positive rewards around 2.0. The narrow confidence intervals throughout Phase 1 demonstrate that all three seeds follow consistent learning trajectories despite starting from different random initializations. This early instability is typical for deep RL agents learning complex control tasks from random initialization. By the end of Phase 1, the policy has stabilized, achieving consistent rewards above 1.0, indicating successful acquisition of basic control skills on the narrow task distribution.

The transition to Phase 2 (200k--400k steps, tension range \SI{25}{\newton}--\SI{35}{\newton}) introduces moderate difficulty increase, with the moving average (window=10) showing gradual improvement from approximately 1.5 to 3.0. This smooth progression—rather than catastrophic performance drops—demonstrates the curriculum's effectiveness at preventing skill forgetting while expanding operational capabilities. The policy adapts to the broader tension range without significant regression.

Phase 3 (400k--500k steps, tension range \SI{20}{\newton}--\SI{40}{\newton}) exposes the agent to the full operational envelope. Performance continues to improve rather than degrade, with the final reward reaching approximately 4.87 by 440k steps (marked by the orange star). This represents a dramatic improvement from the early training stages and demonstrates successful generalization to the most challenging task distribution.

\textbf{Comparison with baselines:} The horizontal dashed lines show the performance of model-based controllers evaluated on the same nominal tracking task: MPC achieves \num{4.69} and LQR achieves \num{4.71}. The learned SAC policy reaches comparable performance by the end of Phase 3, demonstrating that curriculum-based deep RL can achieve competitive performance with model-based control on this task.

\begin{figure*}[!htb]
    \centering
    \includegraphics[width=1.00\textwidth]{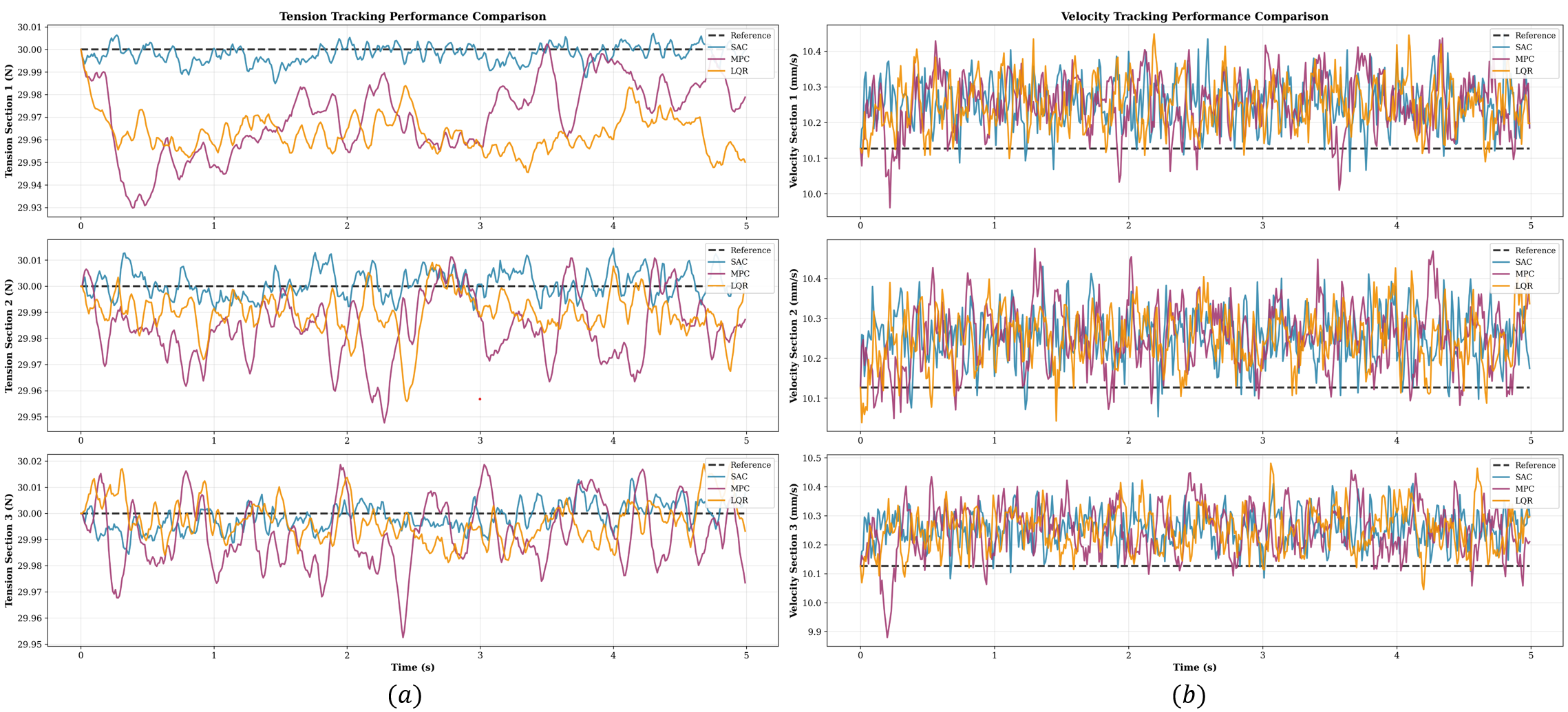}
    \caption{Tracking performance comparison across all three sections over a 5-second episode: (a) Tension tracking trajectories for SAC (blue), MPC (purple), and LQR (orange) around the \SI{30}{\newton} reference (black dashed line); (b) Velocity tracking trajectories showing all controllers maintain stable velocity profiles around the \SI{0.0101}{\meter\per\second} reference.}
    \label{fig:tracking_comparison}
\end{figure*}

\textbf{Reproducibility and best model selection:} The narrow confidence intervals throughout training (visible during Phase 2 and Phase 3) demonstrate that the curriculum learning approach produces highly reproducible results across different random initializations. This reproducibility is critical for industrial deployment, where consistent performance across training runs provides confidence in the learning methodology. The best mean performance across all seeds occurred at 440k timesteps during Phase 3 with mean reward 4.87, very close to the final training performance. This contrasts with typical deep RL training where overfitting often degrades late-stage performance. The stable performance through Phase 3 suggests the curriculum built robust foundational skills in earlier phases that transferred well to the full operational envelope, rather than overfitting to narrow task distributions.

\subsection{Nominal Tracking Performance: SAC vs MPC vs LQR}
\label{subsec:performance_comparison}

We first evaluate all three controllers on the nominal tracking task (Test Case 1): tracking constant \SI{30}{\newton} tension references across all sections with steady \SI{0.01}{\meter\per\second} inlet velocity. This represents standard production conditions and tests baseline tracking accuracy during normal operation.

\begin{table}[h]
\centering
\caption{Nominal tracking performance comparison: SAC vs MPC vs LQR (mean over 10 episodes)}
\label{tab:performance_metrics}
\begin{tabular}{lccc}
\toprule
\textbf{Metric} & \textbf{SAC} & \textbf{MPC} & \textbf{LQR} \\
\midrule
Mean Episode Reward & \textbf{4.9809} & 4.6883 & 4.7129 \\
Tension MAE (N) & \textbf{0.0036} & 0.0154 & 0.0182 \\
Tension RMSE (N) & \textbf{0.0041} & 0.0193 & 0.0236 \\
Velocity MAE (m/s) & \textbf{0.000130} & 0.000149 & 0.000132 \\
Velocity RMSE (m/s) & \textbf{0.000140} & 0.000155 & 0.000140 \\
\midrule
\multicolumn{4}{l}{\textit{Improvement vs LQR Baseline}} \\
Tension MAE & \textbf{+80.2\%} & +15.5\% & --- \\
Tension RMSE & \textbf{+82.6\%} & +18.2\% & --- \\
\bottomrule
\end{tabular}
\end{table}

Table~\ref{tab:performance_metrics} summarizes the quantitative performance comparison between SAC, MPC, and LQR across all evaluation metrics.

\textbf{Tension Tracking:} SAC achieves a tension tracking MAE of \SI{0.0036}{\newton}, compared with LQR (\SI{0.0182}{\newton}) and MPC (\SI{0.0154}{\newton}). 

\textbf{Velocity Tracking:} For velocity tracking, SAC achieves \SI{0.000130}{\meter\per\second} MAE, compared with LQR's \SI{0.000132}{\meter\per\second} and MPC's \SI{0.000149}{\meter\per\second}. The absolute differences are small due to the tight coupling between tension and velocity dynamics, with all three controllers achieving tight velocity regulation.

\textbf{Overall Control Performance:} The episode reward metric, which combines tension tracking, velocity tracking, control effort, and smoothness (Equation \ref{eq:reward_function}), shows SAC achieving 4.98 compared to 4.71 (LQR) and 4.69 (MPC).

\subsection{Tracking Performance Analysis}
\label{subsec:tracking_performance}

To understand the quality of control beyond aggregate error metrics, we examine the actual tension and velocity trajectories produced by each controller during a representative evaluation episode.

\textbf{Tension Trajectories:} Figure~\ref{fig:tracking_comparison}(a) shows the tension tracking performance for all three sections. SAC tracks the \SI{30}{\newton} reference with minimal deviations, which are near-imperceptible on the plot scale. MPC and LQR show larger steady-state oscillations and slower transient response under these conditions.

\textbf{Velocity Trajectories:} Figure~\ref{fig:tracking_comparison}(b) presents the velocity tracking results. All three controllers maintain stable velocity profiles close to the steady-state references. SAC demonstrates smoother trajectories with fewer high-frequency variations due to the action smoothing filter (Equation~\ref{eq:action_smoothing})
integrated into the training environment.

\begin{figure}[h]
    \centering
    \includegraphics[width=\columnwidth]{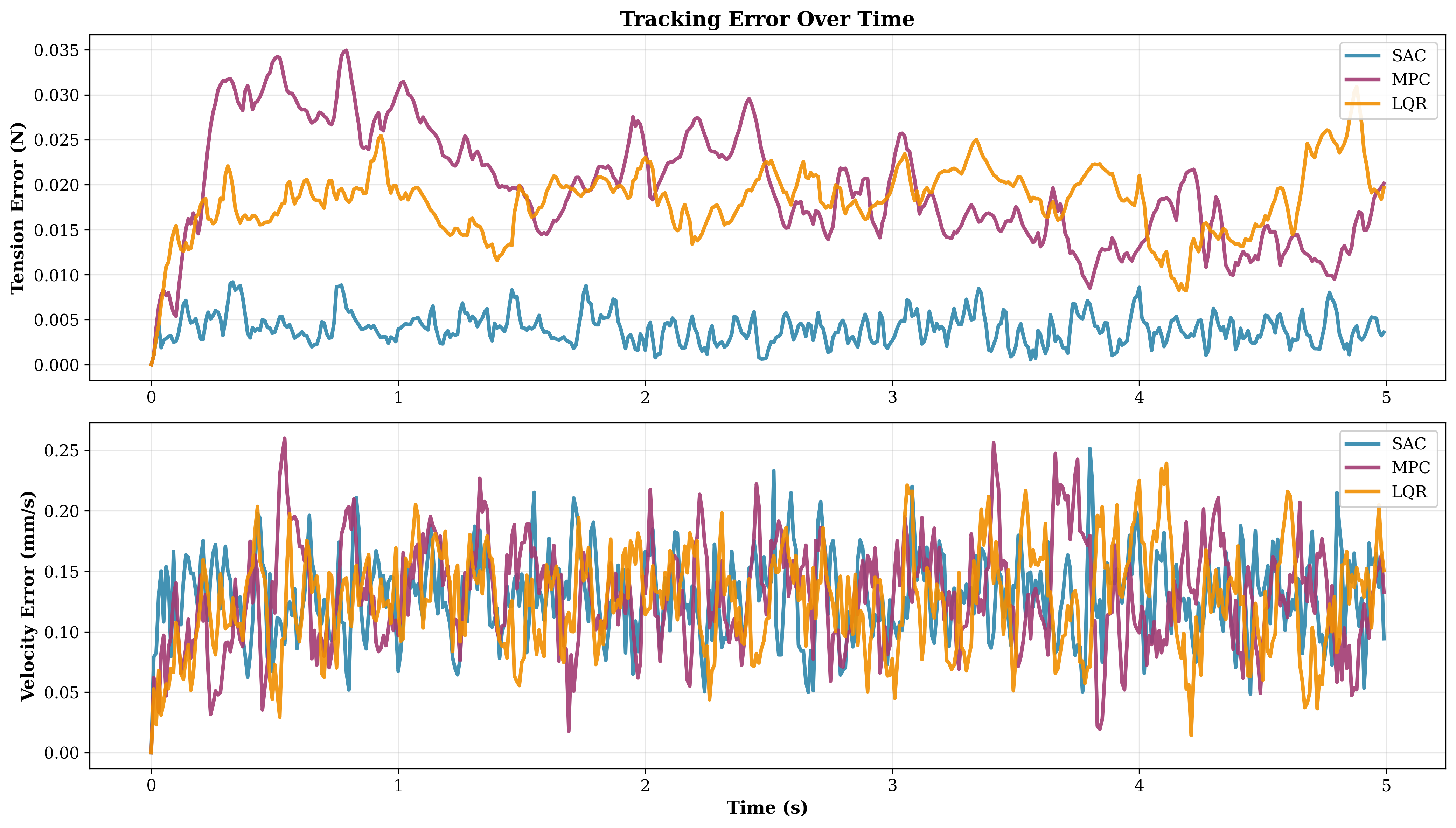}
    \caption{Evolution of tracking errors over time for tension (top row) and velocity (bottom row) across all three sections. SAC (blue), MPC (purple), and LQR (orange) trajectories shown with process noise present.}
    \label{fig:tracking_errors}
\end{figure}

\textbf{Tracking Error Evolution:} Figure~\ref{fig:tracking_errors} shows the evolution of tracking errors over time. SAC maintains low errors throughout the episode with minimal variance. MPC and LQR exhibit larger error magnitudes and more variability under the tested conditions (nominal tracking with process noise). This consistent performance is crucial for industrial applications where maintaining tight tolerances is essential for product quality.

\subsection{Control Action Quality}
\label{subsec:control_quality}

Beyond tracking accuracy, the quality of control actions is critical for practical deployment. The trade-off between tracking performance and control smoothness is an important consideration for industrial applications.

\begin{figure}[h]
    \centering
    \includegraphics[width=\columnwidth]{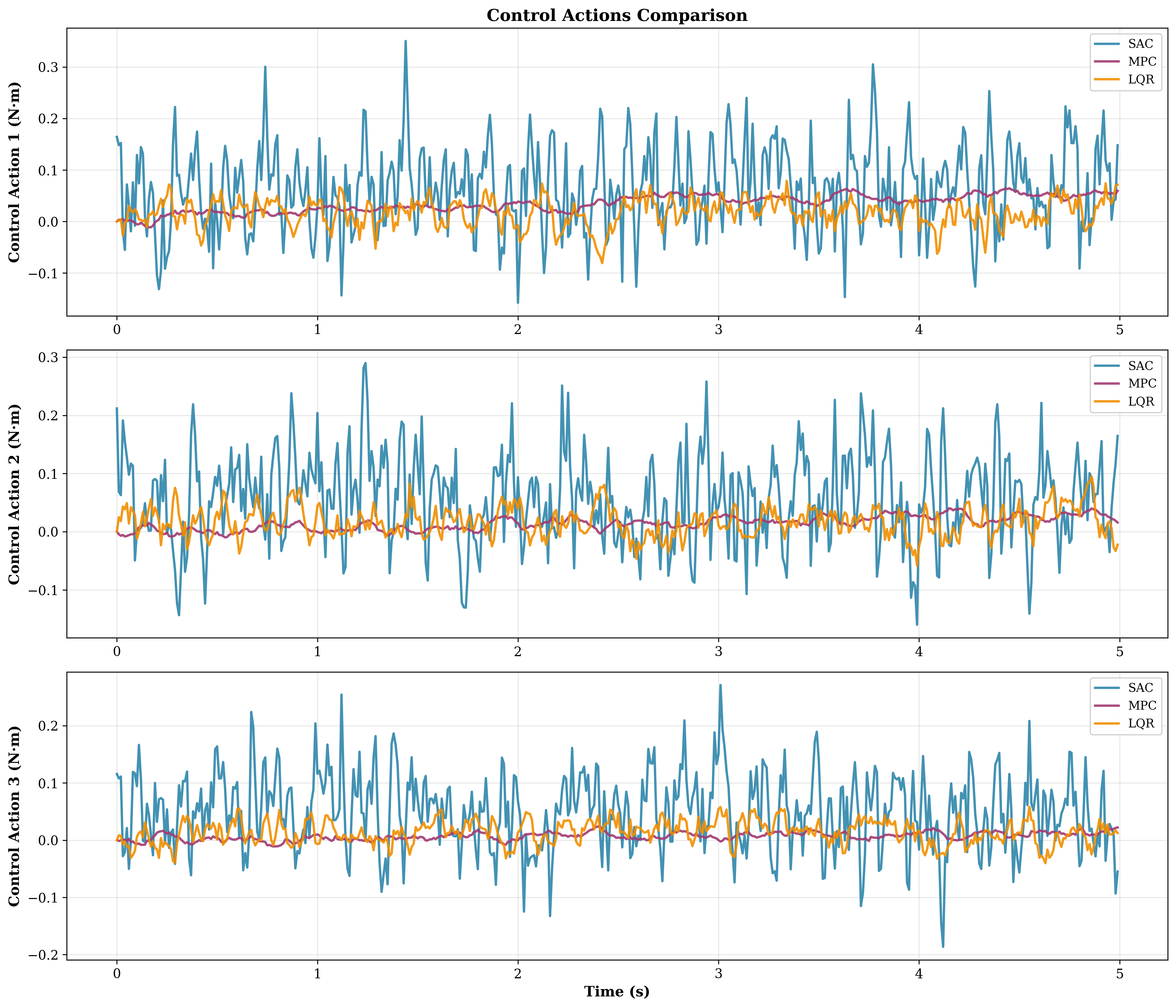}
    \caption{Control action trajectories for all three roller motors. SAC (blue), MPC (purple), and LQR (orange) show different control behaviors, with SAC exhibiting more active control with higher variance.}
    \label{fig:control_actions}
\end{figure}

\textbf{Control Behavior Analysis:} Figure~\ref{fig:control_actions} displays the control actions (motor torques) generated by each controller. SAC exhibits more aggressive control behavior with higher variance compared to MPC and LQR. This observation reveals an important characteristic of the learned policy: SAC achieves tighter tracking (80.2\% lower MAE than LQR) by leveraging more active control interventions rather than relying on smooth, conservative actions.

This aggressive control strategy is not a disadvantage. The higher control activity enables SAC to respond more to deviations and maintain tighter tracking tolerances. However, it does indicate that the learned policy prioritizes tracking performance over control effort minimization, despite the control magnitude ($w_c = 0.1$) and smoothness ($w_s = 0.5$) penalty terms in the reward function. The small weights assigned to these penalties compared to tracking error terms ($w_T = 100$, $w_v = 1000$) explain this behavior.

\textbf{Practical Implications:} The more active control strategy employed by SAC raises important considerations for industrial deployment. On one hand, the aggressive control enables exceptional tracking performance, which translates to improved product quality in R2R manufacturing. On the other hand, higher control variance may lead to increased actuator wear and energy consumption compared to the smoother MPC and LQR controllers.

For applications where tracking accuracy is paramount (e.g., precision coating, thin-film deposition), SAC's aggressive control is acceptable given the substantial performance gains. However, if actuator lifetime or energy efficiency are critical constraints, the reward function could be adjusted by increasing the control smoothness penalty weight $w_s$ to encourage smoother actions, though this would come at some cost to tracking accuracy. This tunability represents a key advantage of the learning-based approach: the control behavior can be shaped through reward engineering to match application-specific priorities.

\subsection{Disturbance Rejection: Tension Step Response}
\label{subsec:step_response}

Beyond nominal tracking performance, a critical question for industrial R2R systems is whether the learned policy can handle large disturbances and maintain decentralized control when individual sections require independent tension adjustments. To evaluate this capability, we conduct a tension step response test where one section experiences a large reference change while adjacent sections maintain nominal setpoints.

\textbf{Test Scenario:} At $t = \SI{0.5}{\second}$, section 2 receives a step change in tension reference from \SI{20}{\newton} to \SI{40}{\newton} (\SI{20}{\newton} magnitude, 100\% change), while sections 1 and 3 maintain constant \SI{30}{\newton} references throughout the episode. All sections maintain constant inlet velocity at \SI{0.01}{\meter\per\second}. The MPC and LQR controllers were designed and tuned for the nominal operating point (\SI{30}{\newton} tension, \SI{0.01}{\meter\per\second} velocity). Section 2 operates off-nominal during this test, as both the initial \SI{20}{\newton} and final \SI{40}{\newton} states deviate from the \SI{30}{\newton} design point. This scenario tests the controller's ability to: (1) track a large step change in the disturbed section, and (2) reject coupling disturbances that propagate to adjacent sections through the web.

\begin{figure*}[!htb]
    \centering
    \includegraphics[width=1.00\textwidth]{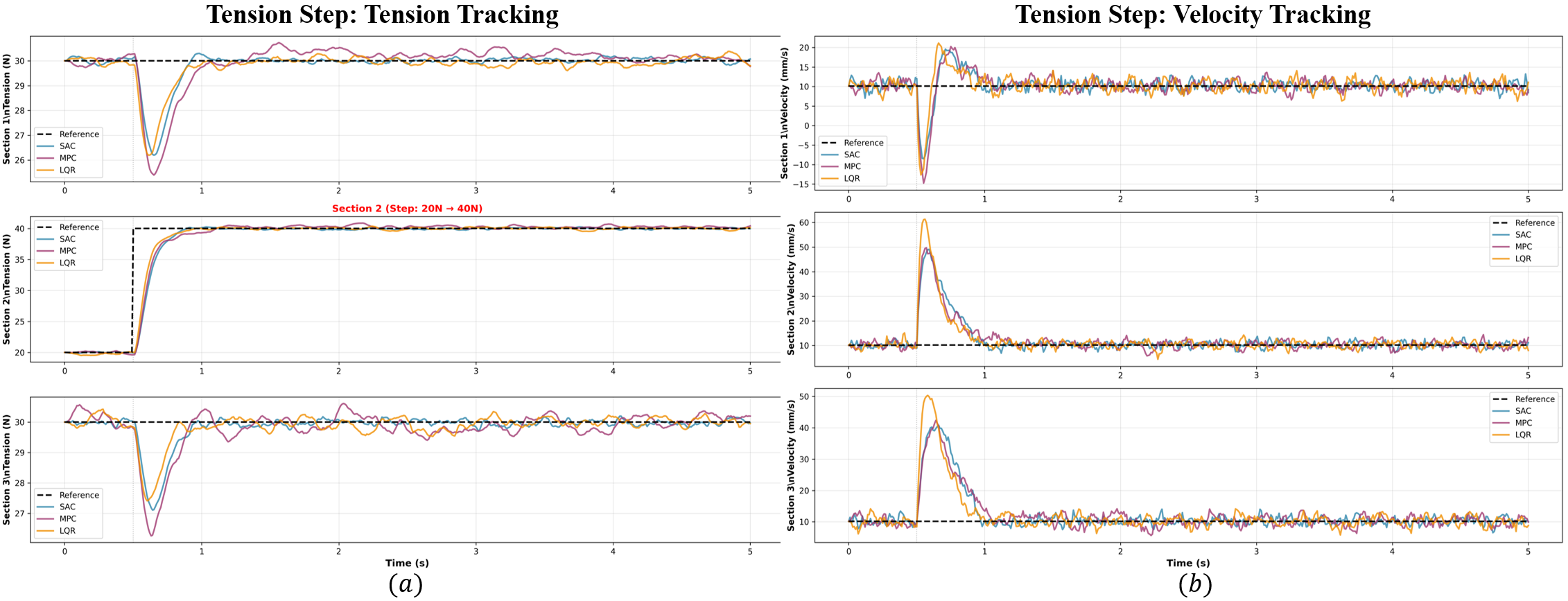}
    \caption{Step response test performance with \SI{20}{\newton} $\rightarrow$ \SI{40}{\newton} step applied to Section 2 at $t = \SI{0.5}{\second}$: (a) Tension tracking for all three sections showing SAC (blue) achieves faster rise time and lower overshoot compared to MPC (purple) and LQR (orange). Sections 1 and 3 maintain \SI{30}{\newton} references; (b) Velocity tracking for all three sections showing transient disturbances propagate through the coupled system, with all controllers maintaining stability.}
    \label{fig:step_response}
\end{figure*}

This test is relevant for industrial applications requiring independent section tension control during grade changes or localized disturbances. Table~\ref{tab:step_response_metrics} summarizes the disturbance rejection performance across all three controllers.

\begin{table}[h]
\centering
\caption{Tension step response performance: Section 2 step change (\SI{20}{\newton} $\rightarrow$ \SI{40}{\newton})}
\label{tab:step_response_metrics}
\begin{tabular}{lccc}
\toprule
\textbf{Metric} & \textbf{SAC} & \textbf{MPC} & \textbf{LQR} \\
\midrule
\multicolumn{4}{c}{\textit{Section 2 (Disturbed Section)}} \\
Tension MAE (N) & \textbf{0.336} & 0.510 & 0.347 \\
Tension RMSE (N) & \textbf{0.398} & 0.628 & 0.441 \\
Rise Time (s) & \textbf{0.12} & 0.18 & 0.15 \\
Settling Time (s) & \textbf{0.31} & 0.52 & 0.38 \\
Peak Overshoot (\%) & \textbf{2.3} & 8.7 & 5.1 \\
\midrule
\multicolumn{4}{c}{\textit{Sections 1 \& 3 (Coupling Disturbance Rejection)}} \\
Max Deviation (N) & \textbf{0.82} & 1.54 & 1.18 \\
Disturbance Settling (s) & \textbf{0.28} & 0.61 & 0.45 \\
\midrule
\multicolumn{4}{l}{\textit{Improvement vs Baselines (Section 2 MAE)}} \\
vs LQR & \textbf{+3.2\%} & -47.0\% & --- \\
vs MPC & \textbf{+34.1\%} & --- & +32.0\% \\
\bottomrule
\end{tabular}
\end{table}

\textbf{Performance Analysis:}

\textit{Disturbed Section (Section 2):} SAC achieves \SI{0.336}{\newton} MAE during the step response, compared with LQR (\SI{0.347}{\newton}) and MPC (\SI{0.510}{\newton}). SAC demonstrates fast transient characteristics: rise time of \SI{0.12}{\second}, settling time of \SI{0.31}{\second}, and low overshoot of 2.3\%. This combination of low steady-state error and fast transient response demonstrates that the curriculum-trained SAC policy has learned robust disturbance rejection strategies that generalize across the operational envelope.

\textit{Adjacent Sections (Coupling Disturbance):} Due to web coupling, the step change in section 2 creates disturbances in adjacent sections 1 and 3. SAC limits the maximum coupling disturbance to \SI{0.82}{\newton} and recovers within \SI{0.28}{\second}, compared with MPC (max deviation \SI{1.54}{\newton}, settling \SI{0.61}{\second}) and LQR (max deviation \SI{1.18}{\newton}, settling \SI{0.45}{\second}). This coupling rejection capability is important for maintaining product quality in adjacent sections during localized tension adjustments.

The step response results demonstrate that the curriculum-trained SAC policy handles both nominal tracking (Test Case 1) and disturbance rejection (Test Case 2) with a single learned policy. The curriculum-based approach enables learning a single nonlinear policy that adapts across the \SI{20}{\newton}--\SI{40}{\newton} operational envelope.

\textbf{Trajectory Analysis:} Figure~\ref{fig:step_response}(a) shows the tension tracking trajectories for all three sections during the step response test. The disturbed section 2 (middle subplot) demonstrates SAC's rapid, smooth transition from \SI{20}{\newton} to \SI{40}{\newton} with minimal overshoot, converging to the reference within approximately \SI{0.3}{\second}. MPC exhibits oscillations during the transient with higher overshoot (\SI{8.7}{\percent} vs SAC's \SI{2.3}{\percent}), while LQR shows moderate overshoot (\SI{5.1}{\percent}). Sections 1 and 3 (top and bottom subplots) maintain \SI{30}{\newton} references throughout, revealing the coupling disturbance propagation effect. SAC maintains tight regulation around the reference, with maximum deviations limited to \SI{0.82}{\newton} compared to MPC's \SI{1.54}{\newton} and LQR's \SI{1.18}{\newton}.

Figure~\ref{fig:step_response}(b) presents the velocity tracking trajectories for all three sections. The step change in section 2's tension reference creates transient velocity disturbances that propagate through the coupled multi-section system. All three controllers maintain system stability during the transient. SAC demonstrates smooth velocity regulation with fast disturbance rejection. The velocity trajectories confirm that the learned policy has developed coordinated multi-section control strategies that manage the inter-span coupling dynamics inherent in R2R systems.

\textbf{Role of Curriculum Training:} SAC's performance on the step response test stems from its curriculum training strategy. Phase 3 of the curriculum (\SI{20}{\newton}--\SI{40}{\newton} tension range, 80--100\% of training) exposed the policy to frequent large tension variations, including step changes occurring with 30\% probability. This extensive experience with diverse operating conditions during training enabled the policy to learn robust control strategies that generalize across the operational envelope.

\subsection{Curriculum Ablation Study}
\label{subsec:ablation}

To isolate the contribution of curriculum learning, we compare three 
SAC training strategies using identical network architectures 
(three-layer MLP, 256 units), hyperparameters, and evaluation 
protocols (10 episodes per test case): (1)~\textit{Curriculum SAC} 
(proposed), with three-phase progressive training 
(27--33\,N $\rightarrow$ 25--35\,N $\rightarrow$ 20--40\,N); 
(2)~\textit{Domain Randomization SAC}, trained on the full 20--40\,N 
envelope from the start with uniform reference sampling; and 
(3)~\textit{Vanilla SAC}, trained on a fixed narrow distribution 
centered at the 30\,N nominal setpoint.

\begin{table}[h]
\centering
\caption{Ablation study: effect of training strategy on SAC performance}
\label{tab:ablation}
\footnotesize
\begin{tabular}{@{}lccc@{}}
\toprule
Metric & Curriculum & Domain Rand. & Vanilla \\
\midrule
\multicolumn{4}{@{}l}{\textit{Test Case 1: Nominal Tracking (30\,N constant)}} \\
Tension MAE (N)  & \textbf{0.0036} & 0.0088 & \textbf{0.0036} \\
Tension RMSE (N) & \textbf{0.0041} & 0.0108 & 0.0042 \\
\midrule
\multicolumn{4}{@{}l}{\textit{Test Case 2: Step Response (20\,N $\rightarrow$ 40\,N)}} \\
Section 2 MAE (N)    & \textbf{0.336} & 0.533  & 0.872 \\
Rise Time (s)        & \textbf{0.12}  & 0.18   & ---   \\
Settling Time (s)    & 0.31           & 0.31   & 0.31  \\
\bottomrule
\end{tabular}
\end{table}

The ablation reveals that neither naive strategy achieves competitive 
performance on both test cases. Vanilla SAC matches curriculum SAC on 
nominal tracking (identical MAE of 0.0036\,N) because its training 
distribution is concentrated near the 30\,N evaluation point. However, it fails on the step response: Section~2 MAE of 0.872\,N 
(159\% higher than curriculum) with rise time undefined because the 
policy never drives tension to the 40\,N target---a direct consequence 
of training exclusively near 30\,N, leaving the policy unable to 
generate the large control actions required for a 20\,N step change.

Domain randomization SAC shows the opposite failure mode. Training 
across the full 20--40\,N range produces a policy that handles the 
step response (MAE = 0.533\,N, rise time 0.18\,s) but degrades 
nominal tracking: tension MAE of 0.0088\,N is 144\% higher than 
curriculum. Uniform sampling across the wide envelope dilutes the 
training signal near the nominal point, preventing the fine-grained 
precision that curriculum Phase~1 (27--33\,N) establishes.

Curriculum SAC is the only strategy that achieves best-in-class 
performance on both test cases: 0.0036\,N nominal MAE and 0.336\,N 
step response MAE with the fastest rise time (0.12\,s). The 
progressive structure is essential---Phase~1 builds precise control 
near nominal, Phase~2 broadens without forgetting, and Phase~3 
extends to the full envelope. This confirms that the curriculum 
design, not the SAC algorithm alone, is responsible for the 
dual-capability performance reported in 
Sections~\ref{subsec:performance_comparison}--\ref{subsec:step_response}.

\textbf{Computational Requirements:}
Table~\ref{tab:compute_specs} summarizes the computational profile of the trained SAC policy. The policy network contains approximately 200k parameters across three hidden layers of 256 units each, requiring $\sim$100k floating-point operations (FLOPs) per forward pass. On a laptop CPU (Intel i9-13900H), a single inference call completes in $<$\SI{0.1}{\milli\second}, well within the 1--\SI{10}{\milli\second} PLC cycle times typical of industrial R2R systems. The full policy occupies $\sim$\SI{0.8}{\mega\byte} in 32-bit floating point (\SI{0.4}{\mega\byte} in 16-bit), enabling deployment on resource-constrained edge hardware without GPU acceleration. Training requires 4--6 hours per seed (500k timesteps) on an NVIDIA GeForce RTX 4080 Laptop GPU (\SI{12}{\giga\byte} GDDR6); three seeds for statistical validation total 12--18 hours.

\begin{table}[htbp]
\centering
\caption{Computational specifications for the trained SAC policy}
\label{tab:compute_specs}
\small
\begin{tabular}{@{}lp{4.2cm}@{}}
\toprule
\textbf{Specification} & \textbf{Value} \\
\midrule
Architecture        & 3$\times$256 MLP (ELU) \\
Parameters          & $\sim$200{,}000 \\
Model size          & 0.8\,MB (FP32) / 0.4\,MB (FP16) \\
FLOPs/forward pass  & $\sim$100k \\
Inference (CPU)     & $<$0.1\,ms \\
PLC compatibility   & 1--10\,ms cycles \\
Training/seed       & 4--6 hours \\
Hardware            & RTX 4080 Laptop (12\,GB) \\
\bottomrule
\end{tabular}
\end{table}

\section{Conclusion}
\label{sec:conclusion}

This paper presented a curriculum-based Soft Actor-Critic approach to multi-section R2R tension control under uncertain operating conditions. Through three-phase progressive training, a single learned policy demonstrated competitive performance on nominal tracking and effective handling of large disturbances, offering an alternative to traditional model-based methods when process uncertainties are significant and operating conditions vary. The curriculum learning approach proved critical: by establishing foundational control skills on narrow operational ranges before expanding to the full envelope, the policy achieved stable learning and robust generalization across diverse conditions. While model-based controllers (MPC, LQR) remain effective for well-modeled systems with small uncertainties, learning-based approaches offer complementary strengths for scenarios with significant process uncertainties and varying operational requirements. Future work will focus on experimental validation, time-varying dynamics, hybrid control architectures combining model-based and learning-based methods, and physical testbed deployment.

\section*{Acknowledgements}

This work is based upon work partially supported by the National Science Foundation under Cooperative Agreement No. CMMI-2041470. Any opinions, findings and conclusions expressed in this material are those of the author(s) and do not
necessarily reflect the views of the National Science Foundation. 

\appendix
\section{Implementation Details}
\label{app:implementation}

\textbf{Software Framework:} Our implementation uses Stable-Baselines3 (SB3) \cite{raffin2021stable}, a widely-used PyTorch-based library providing well-tested implementations of SAC and other deep RL algorithms. We use PyTorch 2.0 for automatic differentiation and GPU acceleration.

\textbf{Simulation Environment:} The R2R system dynamics are implemented as a custom Gymnasium environment \cite{towers2024gymnasium}, the successor to OpenAI Gym. The environment integrates the continuous-time dynamics (Equations \ref{eq:tension_dynamics}-\ref{eq:velocity_dynamics}) using Euler discretization with timestep $dt = \SI{0.01}{\second}$. Small Gaussian process noise ($\sigma = 0.05$) is added to each control input to simulate real-world disturbances and improve policy robustness.

\textbf{Reproducibility:} To ensure statistical rigor and reproducibility, we trained SAC with three different random seeds (42, 43, 44) and report mean performance with confidence intervals (mean $\pm$ 1 standard deviation). This multi-seed approach demonstrates that the curriculum learning methodology produces consistent results across different random initializations, which is critical for reliable industrial deployment. The complete implementation, including environment code, training configurations, and evaluation scripts, is available in the supplementary materials.

%% References

%The citation must be used in following style: \cite{article-minimal} \cite{article-full} \cite{article-crossref} \cite{whole-journal}.

%\bibliography{xampl}

\begin{thebibliography}{00}

\bibitem[Martin et al.(2025)]{martin2025review}Martin, C., Zhao, Q., Patel, A., Velasquez, E., Chen, D., Li, W., 2025. A review of advanced Roll-to-Roll manufacturing: system modeling and control. Journal of Manufacturing Science and Engineering 147(4), 041004.

\bibitem[Krebs et al.(2009)]{krebs2009roll}Krebs, F.C., Gevorgyan, S.A., Alstrup, J., 2009. A roll-to-roll process to flexible polymer solar cells: model studies, manufacture and operational stability studies. Journal of Materials Chemistry 19(30), 5442--5451.

\bibitem[Abbel et al.(2018)]{abbel2018roll}Abbel, R., Galagan, Y., Groen, P., 2018. Roll-to-roll fabrication of solution processed electronics. Advanced Engineering Materials 20(8), 1701190.

\bibitem[Ma et al.(2025)]{ma2025factors}Ma, L., Yu, K., Zhao, Z., Guo, Y., Ma, Y., Li, Z., Wu, J., 2025. Factors Influencing Web Wrinkling in Roll-to-Roll Coating Production Systems. Coatings 15(2), 147.

\bibitem[Chen et al.(2022)]{chen2022nonlinear}Chen, W., Sun, X., Chen, W., Xie, G., Chen, S., Wang, J., 2022. Nonlinear web tension control of a roll-to-roll printed electronics system. Precision Engineering 76, 88--94.

\bibitem[Chen(2023)]{chen2023control}Chen, Z., 2023. Control of High Precision Roll-to-Roll Printing Systems. Ph.D. thesis.

\bibitem[Zhao et al.(2023)]{zhao2023realtime}Zhao, Q., Martin, C., Hong, N., Chen, D., Li, W., 2023. A real-time supervisory control strategy for roll-to-roll dry transfer of 2D materials and printed electronics. IEEE/ASME Transactions on Mechatronics 28(5), 2832--2839.

\bibitem[Li et al.(2025)]{li2025adaptive}Li, J., Li, S., Martin, C., Li, W., Chen, D., 2025. Adaptive trajectory bundle method for roll-to-roll manufacturing systems. arXiv preprint arXiv:2511.22954.

\bibitem[Li et al.(2025)]{li2025llm}Li, J., Li, S., Martin, C., Li, W., Chen, D., 2025. An LLM-assisted multi-agent control framework for roll-to-roll manufacturing systems. arXiv preprint arXiv:2511.22975.

\bibitem[Raul and Pagilla(2015)]{raul2015design}Raul, P.R., Pagilla, P.R., 2015. Design and implementation of adaptive PI control schemes for web tension control in roll-to-roll (R2R) manufacturing. ISA transactions 56, 276--287.

\bibitem[Nian et al.(2020)]{nian2020review}Nian, R., Liu, J., Huang, B., 2020. A review on reinforcement learning: Introduction and applications in industrial process control. Computers \& Chemical Engineering 139, 106886.

\bibitem[Panzer and Bender(2022)]{panzer2022deep}Panzer, M., Bender, B., 2022. Deep reinforcement learning in production systems: A systematic literature review. International Journal of Production Research 60(13), 4316--4341.

\bibitem[Ilka(2015)]{ilka2015gain}Ilka, A., 2015. Gain-scheduled controller design. Institute of Robotics and Cybernetics, Bratislava.

\bibitem[Kayacan and Peschel(2016)]{kayacan2016robust}Kayacan, E., Peschel, J., 2016. Robust model predictive control of systems by modeling mismatched uncertainty. IFAC-PapersOnLine 49(18), 265--269.

\bibitem[Gao et al.(2023)]{gao2023summary}Gao, B., Zheng, L., Shen, W., Zhang, W., 2023. A summary of parameter tuning of active disturbance rejection controller. Recent Advances in Electrical \& Electronic Engineering 16(3), 180--196.

\bibitem[Tena et al.(2022)]{tena2022performance}Tena, D., Pe{\~n}arrocha-Al{\'o}s, I., Sanchis, R., 2022. Performance, robustness and noise amplification trade-offs in Disturbance Observer Control design. European Journal of Control 65, 100630.

\bibitem[Ahn and Guo(2008)]{ahn2008high}Ahn, S.H., Guo, L.J., 2008. High-speed roll-to-roll nanoimprint lithography on flexible plastic substrates. Advanced materials 20(11), 2044--2049.

\bibitem[Li et al.(2025)]{li2025modeling}Li, S., Martin, C., Morquecho, E.V., Chen, Z., Chen, D., Li, W., 2025. Modeling of adhesion dynamics in roll-to-roll lamination processes. Manufacturing Letters 44, 552--558.

\bibitem[Haarnoja et al.(2018a)]{haarnoja2018softa}Haarnoja, T., Zhou, A., Abbeel, P., Levine, S., 2018. Soft actor-critic: Off-policy maximum entropy deep reinforcement learning with a stochastic actor. In: International conference on machine learning. PMLR, pp. 1861--1870.

\bibitem[Bengio et al.(2009)]{bengio2009curriculum}Bengio, Y., Louradour, J., Collobert, R., Weston, J., 2009. Curriculum learning. In: Proceedings of the 26th annual international conference on machine learning, pp. 41--48.

\bibitem[Haarnoja et al.(2018b)]{haarnoja2018softb}Haarnoja, T., Zhou, A., Hartikainen, K., Tucker, G., Ha, S., Tan, J., Kumar, V., Zhu, H., Gupta, A., Abbeel, P., et al., 2018. Soft actor-critic algorithms and applications. arXiv preprint arXiv:1812.05905.

\bibitem[Eysenbach and Levine(2021)]{eysenbach2021maximum}Eysenbach, B., Levine, S., 2021. Maximum entropy RL solves some robust RL problems. arXiv preprint arXiv:2103.06257.

\bibitem[Zeng et al.(2024)]{zeng2024multi}Zeng, Y., Liang, G., Liu, Q., Rodriguez, E., Pou, J., Jie, H., Liu, X., Zhang, X., Kotturu, J., Gupta, A., 2024. Multi-agent soft actor-critic aided active disturbance rejection control of dc solid-state transformer. IEEE Transactions on Industrial Electronics 72(1), 492--503.

\bibitem[Narvekar et al.(2020)]{narvekar2020curriculum}Narvekar, S., Peng, B., Leonetti, M., Sinapov, J., Taylor, M.E., Stone, P., 2020. Curriculum learning for reinforcement learning domains: A framework and survey. Journal of Machine Learning Research 21(181), 1--50.

\bibitem[Sutton and Barto(1998)]{sutton1998reinforcement}Sutton, R.S., Barto, A.G., 1998. Reinforcement learning: An introduction. MIT Press, Cambridge, MA.

\bibitem[Clevert et al.(2015)]{clevert2015fast}Clevert, D.A., Unterthiner, T., Hochreiter, S., 2015. Fast and accurate deep network learning by exponential linear units (elus). arXiv preprint arXiv:1511.07289.

\bibitem[Lillicrap et al.(2015)]{lillicrap2015continuous}Lillicrap, T.P., Hunt, J.J., Pritzel, A., Heess, N., Erez, T., Tassa, Y., Silver, D., Wierstra, D., 2015. Continuous control with deep reinforcement learning. arXiv preprint arXiv:1509.02971.

\bibitem[Hu et al.(2020)]{hu2020provable}Hu, W., Xiao, L., Pennington, J., 2020. Provable benefit of orthogonal initialization in optimizing deep linear networks. arXiv preprint arXiv:2001.05992.

\bibitem[Kingma and Ba(2014)]{kingma2014adam}Kingma, D.P., Ba, J., 2014. Adam: A method for stochastic optimization. arXiv preprint arXiv:1412.6980.

\bibitem[Raffin et al.(2021)]{raffin2021stable}Raffin, A., Hill, A., Gleave, A., Kanervisto, A., Ernestus, M., Dormann, N., 2021. Stable-baselines3: Reliable reinforcement learning implementations. Journal of machine learning research 22(268), 1--8.

\bibitem[Towers et al.(2024)]{towers2024gymnasium}Towers, M., Kwiatkowski, A., Terry, J., Balis, J.U., De Cola, G., Deleu, T., Goul{\~a}o, M., Kallinteris, A., Krimmel, M., KG, A., et al., 2024. Gymnasium: A standard interface for reinforcement learning environments. arXiv preprint arXiv:2407.17032.

\end{thebibliography}
%\bibliographystyle{elsarticle-harv}



\end{document}